\input psfig
\documentstyle[preprint,aps]{revtex}
\begin{document}

\draft
\title{
Quantum Hall Fluids on the Haldane Sphere:\\ A Diffusion Monte Carlo
Study}
\author{V. Melik-Alaverdian and N. E. Bonesteel}
\address{
National High Magnetic Field Laboratory and Department of Physics,
Florida State University, Tallahassee, FL 32306-4005}
\author{G. Ortiz}
\address{
Theoretical Division, Los Alamos National Laboratory, P.O. Box 1663, 
Los Alamos, NM 87545}

\date{\today}
\maketitle

\begin{abstract}
A generalized diffusion Monte Carlo method for solving the many-body
Schr\"odinger equation on curved manifolds is introduced and used to
perform a `fixed-phase' simulation of the fractional quantum Hall
effect on the Haldane sphere. This new method is used to study the
effect of Landau level mixing on the $\nu=1/3$ energy gap and the
relative stability of spin-polarized and spin-reversed quasielectron
excitations.
\end{abstract}


%\begin{multicols}{2}

\columnseprule 0pt

\narrowtext

Much of the modern understanding of the fractional quantum Hall effect
(FQHE) is based on the observation that in two dimensions the quantum
statistics of identical particles can be changed by performing a
singular gauge transformation \cite{sgt}.  Such a transformation can
be used, for example, to map the equations describing an ideal two
dimensional electron gas in a transverse magnetic field at Landau
level filling factor $\nu=1/q$, where $q$ is an odd integer, into
those describing a system of `composite bosons' moving in zero
effective magnetic field, interacting via both Coulomb and
`Chern-Simons' interactions.  This transformation, from fermions to
bosons, is the basis of the successful Chern-Simons-Landau-Ginzburg
phenomenology of the FQHE \cite{sczhang}.  Its existence also suggests
the possibility of numerically simulating the FQHE at $\nu=1/q$ using
the composite boson description.

Recently, Ortiz, Ceperley, and Martin (OCM) \cite{ortiz} introduced
the `fixed-phase' diffusion Monte Carlo (DMC) method for simulating
non-time-reversal symmetric systems with complex-valued
eigenfunctions.  OCM applied this method to the FQHE ground state at
$\nu=1/q$, using the torus geometry, and fixing the phase of the wave
function with Laughlin's trial wave function.  The resulting effective
bosonic problem corresponded precisely to the composite boson
description, with the additional approximation that those terms in the
transformed Hamiltonian leading to fluctuations of the phase of the
wave function were ignored.  This effective bosonic problem was then
solved by standard DMC techniques \cite{reynolds}, and the results
used to study the effect of Landau level mixing (LLM) on the FQHE
ground state.  However, OCM did not consider either excited states or
geometries other than the torus.

In this Letter we present the results of a `fixed-phase' DMC study of
the FQHE at $\nu=1/3$ using the spherical geometry introduced by
Haldane \cite{haldane}.  In order to go from the torus to the sphere
we introduce a generalized DMC method for solving the many-body
Schr\"odinger equation on curved manifolds.  One motivation for this
work is that the `Haldane sphere' is arguably the most convenient
geometry for numerical study of the FQHE, and we believe that our
generalization of the fixed-phase DMC method to this geometry will be
useful for many future calculations.  As an example of the application
of this new method we have calculated the effect of LLM on the FQHE
transport gap, {\it i.e.}, the energy gap for creating a fractionally
charged quasielectron -- quasihole pair with infinite separation.
Results have been obtained for both a spin-polarized and spin-reversed
quasielectron.  Previous calculations of the crossover magnetic field
below which the transport gap is set by the spin-reversed excitation
have ignored LLM \cite{chak,morf}.  The present work includes these
effects for the first time.

An ideal two dimensional electron gas, with effective mass $m^*$,
carrier density $n$ and dielectric constant $\epsilon$, placed in a
transverse magnetic field $B$ is characterized by three length scales
--- the effective Bohr radius $a_B = \epsilon \hbar^2/m^* e^2$, the
magnetic length $l_0 = \sqrt{\hbar c/eB}$, and the mean interparticle
spacing $d = 1/\sqrt{\pi n}$.  These length scales can be combined to
form two independent dimensionless ratios, the filling factor $\nu = 2
l_0^2/d^2$, and the electron gas parameter $r_s = d/a_B$. The degree
of LLM is characterized by the ratio of the typical Coulomb energy to
the cyclotron energy $(e^2/\epsilon d)/\hbar\omega_c = r_s \nu/2$
where $\omega_c = eB/m^*c$.  Thus, for fixed $\nu$, $r_s$ provides a
useful measure of the importance of LLM.  Because $r_s \propto
m^*/\sqrt{B}$ LLM can be increased either by decreasing $B$ or
increasing $m^*$.  For example, in two dimensional GaAs/AlGaAs systems
with typical carrier densities, for $n$-type systems $m^*
\simeq 0.07$ and $r_s \sim 2$ while for $p$-type systems $m^* \simeq
0.38$ and $r_s \sim 10$.

In the spherical geometry electrons are confined to the surface of a
sphere of radius $R$ with a magnetic monopole at its center.  Let $N$
denote the number of electrons and $2S$ denote the number of flux
quanta piercing the surface of the sphere.  The field strength is then
$B = S\hbar c/e R^2$ and, for $\nu=1/q$, $2S = q(N-1)$.  If electron
positions are given in stereographic coordinates, ${\bf r} = (x,y) =
(\cos\phi,\sin\phi) \tan \theta/2$ where $\theta$ and $\phi$ are the
usual spherical angles, then the Hamiltonian is, in appropriately
scaled atomic units ($\hbar = m^* = e^2 = 1$),
\begin{eqnarray}
H =  \frac{1}{2}
\sum_i D({\bf r}_i)
(-i{\bf \nabla}_i + {\bf A}({\bf r}_i))^2 + \frac{1}{\epsilon}
\sum_{i<j} \frac{1}{\sqrt{r_{ij}^2+\beta^2}} \ ,
\label{ham1}
\end{eqnarray}
where $D({\bf r}_i) = (1+r_i^2)^2/4R^2$, $r_{ij}$ is the chord
distance on the sphere, and $\beta$ is a parameter which models the
finite thickness of the 2DEG \cite{fczhang}.  We work in the Wu-Yang
\cite{wu} gauge for which ${\bf A}({\bf r}_i) = 2R^2 B (-y_i, x_i)/(1+r_i^2)$.

In what follows we have used three trial wave functions to implement
the fixed-phase approximation ($z = x+iy$ is the complex stereographic
coordinate):

{\it (i) Ground state.---} In the Wu-Yang gauge the spherical analog
of the Laughlin wave function for $\nu=1/q$ is \cite{haldane}
\begin{equation}
\psi_{\rm GS} = \prod_k(1+|z_k|^2)^{-S}
\prod_{i<j}(z_i-z_j)^q \ .
\end{equation}

{\it (ii) Spin-polarized excited state.---} For this state we have
used the following wave function, constructed using Jain's composite
fermion approach \cite{jain}, describing an excitation with a charge
$e/q$ quasielectron at the top of the sphere and a charge $-e/q$
quasihole at the bottom of the sphere,
\begin{eqnarray}
\psi_{\rm SP} = \prod_k(1+|z_k|^2)^{-S} 
\prod_{i<j}(z_i-z_j)^{q-1} \left|   
\begin{array}{ccccc}
1 & z_1 & ... & z_1^{N-2} & \sum_{i\ne 1} \frac{1}{z_1-z_i} \\
\vdots & \vdots  &  &\vdots &\vdots  \\
1 & z_N & ... & z_N^{N-2} & \sum_{i\ne N} \frac{1}{z_N-z_i} \\
\end{array}\right| \ .
\end{eqnarray}

{\it (iii) Spin-reversed excited state.---} This state is similar to
$\psi_{\rm SP}$ except the quasielectron has a reversed spin.  If
$z_1$ denotes the coordinate of the down spin electron then this wave
function is
\cite{morf}
\begin{equation}
\psi_{\rm SR} = \prod_k(1+|z_k|^2)^{-S}
\prod_{l\ne 1} (z_l - z_1)^{-1}
\prod_{i<j}(z_i-z_j)^q \ .
\end{equation}

The fixed-phase approximation is carried out as follows.  First, the
relevant trial function is written as $\psi_{\rm T}({\cal R}) = |
\psi_{\rm T}({\cal R})| \ \exp[i \phi_{\rm T}({\cal R})]$, where
${\cal R} = ( {\bf r}_1, {\bf r}_2, \cdots, {\bf r}_N)$.  The overall
phase $\phi_{\rm T}({\cal R})$ is then used to perform a singular
gauge transformation, $\hat H = \exp[-i \phi_{\rm T}({\cal R})] \ H \
\exp[i \phi_{\rm T}({\cal R})] = H_{\rm R} + i H_{\rm I}$, where
\begin{eqnarray}
H_{\rm R} &=& -\frac{1}{2} \sum_i D({\bf r}_i) (\nabla^2_i -
\tilde {\bf A}^2({\bf r}_i) ) + \frac{1}{\epsilon}\sum_{i<j} 
\frac{1}{\sqrt{r_{ij}+\beta^2}}, \\  
H_{\rm I} &=& - \frac{1}{2} \sum_i D( {\bf r}_i) 
\left({\bf \nabla}_i \cdot \tilde {\bf A}({\bf r}_i) 
+\tilde {\bf A}( {\bf r}_i) 
\cdot {\bf \nabla}_i  \right) \ ,
\end{eqnarray}
and $\tilde {\bf A}({\bf r}_i) = {\bf A}({\bf r}_i) + {\bf \nabla}_i
\phi_{\rm T}({\cal R})$.  As shown by OCM, the bosonic ground state of
$H_{\rm R}$ is the lowest energy state with the same phase as the
trial function \cite{ortiz}.  The DMC method can then be applied to
the imaginary time Schr\"odinger equation $H_{\rm R} \psi({\cal R},t)
= - \frac{\partial}{\partial t}\psi({\cal R},t)$, the solution of
which, in the limit $t \rightarrow \infty$, converges to the ground
state of $H_{\rm R}$.

The dependence of $D({\bf r})$ on position in (\ref{ham1}) is a
consequence of the finite curvature of the surface of the sphere, for
which the metric tensor, in stereographic coordinates, is
$g_{\alpha\beta}({\bf r}) = D({\bf r})^{-1} \delta_{\alpha\beta}$.
Below we introduce our generalized DMC method for simulating the
many-body Schr\"odinger equation on such a curved manifold.  Note that
in two dimensions it is always possible to choose coordinates for
which $g_{\alpha\beta} = f({\bf r}) \delta_{\alpha\beta}$, {\it i.e.},
to work in the so-called `conformal gauge', and so the generalized DMC
method introduced below can, in principle, be applied to {\it any}
curved two-dimensional manifold, not just the sphere.

The central modification of the DMC method required to treat the
present problem is to replace the usual importance-sampled
distribution function $P({\cal R},t) = \psi({\cal R},t) |\psi_{\rm
T}({\cal R})|$ \cite{reynolds} with
\begin{eqnarray}
\widetilde P({\cal R},t) = \psi({\cal R},t) |\psi_T({\cal R})| 
\prod_{i=1}^N  \frac{1}{D({\bf r}_i)} \ .
\end{eqnarray}
This has two important consequences.  First, because the differential
area element on the sphere is $dA = d^2r/D({\bf r})$ the expectation
value of the ground state energy is simply
\begin{eqnarray}
\langle  H \rangle = 
\frac{\int \widetilde P({\cal R},t\rightarrow \infty) E_L({\cal R}) d{\cal R}}
{\int \widetilde P({\cal R},t\rightarrow \infty) d{\cal R}} \ .
\end{eqnarray}
Second, the differential equation satisfied by $\widetilde P({\cal
R},t)$ is,
\begin{eqnarray}
- \frac{\partial}{\partial t}\widetilde P({\cal R} ,t) =
\sum_{i=1}^N \left[ -\frac{1}{2}
\nabla_i^2 (D({\bf r}_i) \widetilde P({\cal R},t)) \right.   
+ {\bf \nabla}_i \cdot (D({\bf r}_i) {\bf F}_i({\cal R}) \widetilde
P({\cal R},t))
\mbox{\huge ]} 
+ (E_L({\cal R})-E_T) \widetilde P({\cal R},t) \ ,
\label{forced}
\end{eqnarray}
where ${\bf F}_i({\cal R}) = {\bf \nabla}_i \ln |\psi_T|$, $E_L({\cal
R}) = H_{\rm R} |\psi_{\rm T}|/|\psi_{\rm T}|$ and $E_T$ is a constant
which must be adjusted in the course of the simulation to be equal to
the ground state energy.  It is worth noting that, except for the
position dependence of $D({\bf r})$, (\ref{forced}) has the same form
as the usual generalized diffusion equation appearing in DMC
simulations \cite{reynolds}.

Equation (\ref{forced}) can be solved numerically by stochastically
iterating the integral equation
\begin{eqnarray}
\widetilde P({\cal R}^\prime,t+\tau) = \int G({\cal R} \rightarrow {\cal
R}^\prime, \tau) \widetilde P({\cal R},t) d \cal R \ ,
\label{integral_equation}
\end{eqnarray}
using the short-time propagator (${\cal O} (\tau^2)$)
\begin{eqnarray}
G({\cal R} \rightarrow {\cal R}^\prime, \tau) = \exp\left[ -\tau
\left( \frac{[ E_L({\cal R}) + E_L({\cal R}^\prime) ]}{2}-E_T \right)
\right] 
 \prod_{i=1}^N G^0_i({\cal R} \rightarrow {\cal R}^\prime,
\tau) \ ,
\label{gf1}
\end{eqnarray}
where
\begin{eqnarray}
G_i^0({\cal R} \rightarrow {\cal R}^\prime, \tau) 
= \frac{1}{2 \pi D({\bf r}_i) \tau}  \exp\left[\frac{-({\bf
r}_i^\prime - {\bf r}_i - D( {\bf r}_i) \tau {\bf F}_i({\cal R}))^2}{2
D({\bf r}_i)\tau}\right]
\label{gf2}
\end{eqnarray}
represents a diffusion and drift process.  In (\ref{gf2}) both $D({\bf
r}_i)$ and $F_i({\cal R})$ are evaluated at the `prepoint' in the
integral equation.  This makes it possible to simulate
(\ref{integral_equation}) in terms of branching random walks by a
straightforward application of the rules given in \cite{reynolds}, and
is a direct consequence of having the spatial derivatives in
(\ref{forced}) sit all the way to the left in each term \cite{future}.
This in turn follows from our modified definition of $\widetilde
P({\cal R},t)$.  Had we used the usual definition, $P({\cal R},t)$, we
would have obtained `quantum corrections' in the propagator \cite{qc}.

The $r_s$ dependence of the ground state energy, and the energies of
the spin-polarized and spin-reversed excited states, have been
calculated by fixing the phase with the trial functions $\psi_{\rm
GS}$, $\psi_{\rm SP}$ and $\psi_{\rm SR}$, respectively, and solving
the resulting bosonic problems using the generalized DMC method
outlined above.  Figure 1 shows the results for the ground state
energy as a function of $r_s$ for $\beta = 0$ and $\nu=1/3$, compared
with the variational Monte Carlo results of Price {\it et al.}
\cite{price,note1}.  This comparison provides an important test of our
generalized DMC method --- the wave functions used in the variational
calculations have the same phase as $\psi_{\rm GS}$ and so must have
higher energies than our fixed-phase DMC results, as is in fact the
case.

The spin-polarized and spin-reversed energy gaps, $\Delta_{\rm SP}$
and $\Delta_{\rm SR}$, obtained by subtracting the ground state energy
from the relevant excited state energies, are plotted vs.  $r_s$ in
Fig.~2.  Results are for $N=20$ and are given for $\beta = 0$ and the
more experimentally relevant case $\beta = 1.5\ l_0$. To reduce finite
size effects we have subtracted $V_0 = - (e/q)^2/2\epsilon R$, the
Coulomb energy of two point charges with charge $\pm e/q$ at the top
and bottom of the sphere, from our results for the gaps \cite{nick}.
The additional Zeeman energy of the spin-reversed excitation is not
included in our definition of $\Delta_{\rm SR}$, and so the crossover
magnetic field, $B_c$, below which the spin-reversed excitation has
lower energy than the spin-polarized excitation, is $B_c =
(\Delta_{\rm SP} - \Delta_{\rm SR})/g\mu_B$, where $\mu_B$ is the Bohr
magneton and $g$ is the effective $g$-factor.  For GaAs ($\epsilon
\simeq 13, g\simeq 0.5$) we find, for $r_s = 2$, $B_c \simeq 14$ T,
while for $r_s = 10$, $B_c \simeq 7$ T.  This reduction of $B_c$ with
increasing $r_s$ reflects the fact that, for $\beta = 0$, LLM has a
significantly stronger effect on $\Delta_{\rm SP}$ than on the
$\Delta_{\rm SR}$, and so tends to stabilize the spin-polarized
excitation.

For the more experimentally relevant case $\beta = 1.5 \ l_0$ the
effect of LLM on $\Delta_{\rm SP}$ and $\Delta_{\rm SR}$ is much
weaker than for $\beta = 0$ and the difference in the two gap
energies, in units of $e^2/\epsilon l_0$, is roughly constant.  Again
using GaAs parameters we find $B_c \simeq 4$ T for $r_s =2$ and $B_c
\simeq 3$ T for $r_s = 10$.  This result for the crossover field
is in reasonable agreement with previous calculations which included
the thickness correction but not LLM \cite{chak,morf}.  The new result
here is that, when thickness is included, $B_c$ is only weakly
dependent on LLM.

Figure 3 shows mixed estimates \cite{reynolds} of the density profiles
of the spin-polarized and spin-reversed excited states at $\nu = 1/3$
as a function of $\theta$ for $r_s=1$ and $r_s=20$.  The results are
for $N=$ 20 and $\beta = 0$.  Note that for $r_s=20$ the quasielectron
and quasihole induce ripples in the density.  This is a consequence of
the increased Wigner crystal-like correlations in the FQHE fluid
induced by LLM.

The effect of LLM on $\Delta_{\rm SP}$ and $\Delta_{\rm SR}$ can be
understood qualitatively using Fig.~3 as follows.  For the
spin-polarized state the quasielectron charge is concentrated in a
ring centered at the top of the sphere.  As $r_s$ increases this ring
of charge spreads out and the dip at $\theta = 0$ deepens.  This
evolution of the quasielectron charge occurs because, as higher Landau
levels are mixed into the wave function, the charge is free to spread
out, lowering its Coulomb energy at the cost of some kinetic energy.
For the spin-reversed quasielectron the charge is spread out more
uniformly and the Coulomb energy is less than that of the
spin-polarized quasielectron. There is therefore less of a tendency
for the spin-reversed quasielectron charge to spread out with
increasing $r_s$, and so this excitation is less affected by LLM.  The
weakening of the effect of LLM on the energy gaps due to the thickness
correction can also be understood along similar lines.  For $\beta =
1.5\ l_0$ the short-range part of the Coulomb interaction is softened,
and the potential energy of both the spin-reversed and spin-polarized
quasielectrons are reduced.  There is therefore less of a tendency for
the charge of these excitations to spread out with increasing $r_s$,
and so, again, the effect of LLM is weakened.

For typical carrier densities the magnetic field at $\nu=1/3$ is
greater than $B_c$ and the transport gap is set by $\Delta_{\rm SP}$.
This gap has been measured in both $n$-type \cite{willet} and $p$-type
\cite{manoharan} GaAs quantum wells, with typical results, for the
highest quality samples, of $\Delta_{\rm n} \simeq 0.05 e^2/\epsilon
l_0$ and $\Delta_{\rm p} \simeq 0.023 e^2/\epsilon l_0$, respectively.
The factor of 2 reduction of the energy gap from $n$-type ($r_s \sim
2$) to $p$-type ($r_s \sim 10$) samples has been attributed to the
increased LLM \cite{manoharan}.  However, our results show that LLM
has only a weak effect on $\Delta_{\rm SP}$ when the thickness effect
is included, in agreement with previous calculations
\cite{price,yoshioka,melik-alaverdian}.  Thus, while our energy gap
for $r_s \simeq 2$ is close to the experimental value, our result for
$r_s \simeq 10$ is off by roughly a factor of $2$.  This discrepancy
between theory and experiment is most likely due to disorder, the
effects of which on the energy gap are still poorly understood.

To summarize, a generalized DMC method for solving the many-body
Schr\"odinger equation on curved manifolds has been introduced and
used to perform a `fixed-phase' simulation of the FQHE on the Haldane
sphere.  The effect of LLM on the $\nu=1/3$ energy gap, and the
relative stability of the spin-polarized and spin-reversed
quasielectron states have been investigated using the new method.  We
believe that the generalization of the `fixed-phase' DMC method to the
spherical geometry presented here will be useful for many future
numerical studies of the FQHE.

The authors would like to thank D. Ceperley and L. Engel for useful
discussions.  This work was supported by the U.S. Department of Energy
under Grant No.\ DE-FG02-97ER45639 and by the National High Magnetic
Field Laboratory.  NEB acknowledges support from the Alfred P. Sloan
foundation, and GO from an Oppenheimer fellowship.

%\newpage

\newpage
\begin{figure}
\centerline{
\psfig{figure=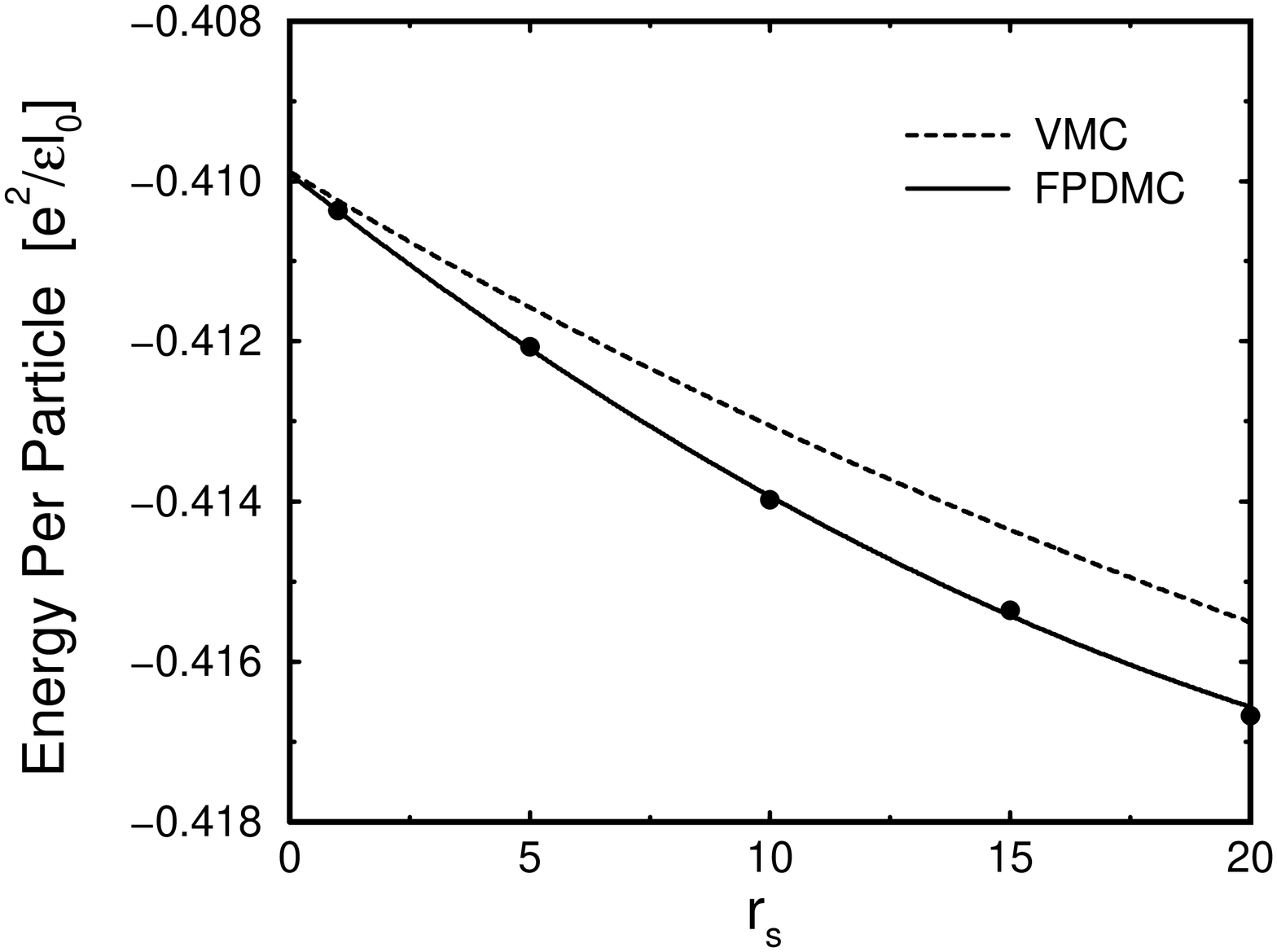,height=3.7in,angle=-90}}
\vskip .5in
\caption{ 
Ground state energy per particle for $\nu=1/3$ as a function of $r_s$.
The dashed line is the variational result of Price {\it et al.}
\protect \cite{price} and the solid line is a least square fit of
second degree polynomial in $r_s$ to our fixed-phase DMC results for
$r_s=$ 1, 5, 10, 15 and 20 (dots).  Statistical errors are smaller
than symbol sizes.}
\label{gapvsl}
\end{figure}

\newpage

\begin{figure}
\centerline{
\psfig{figure=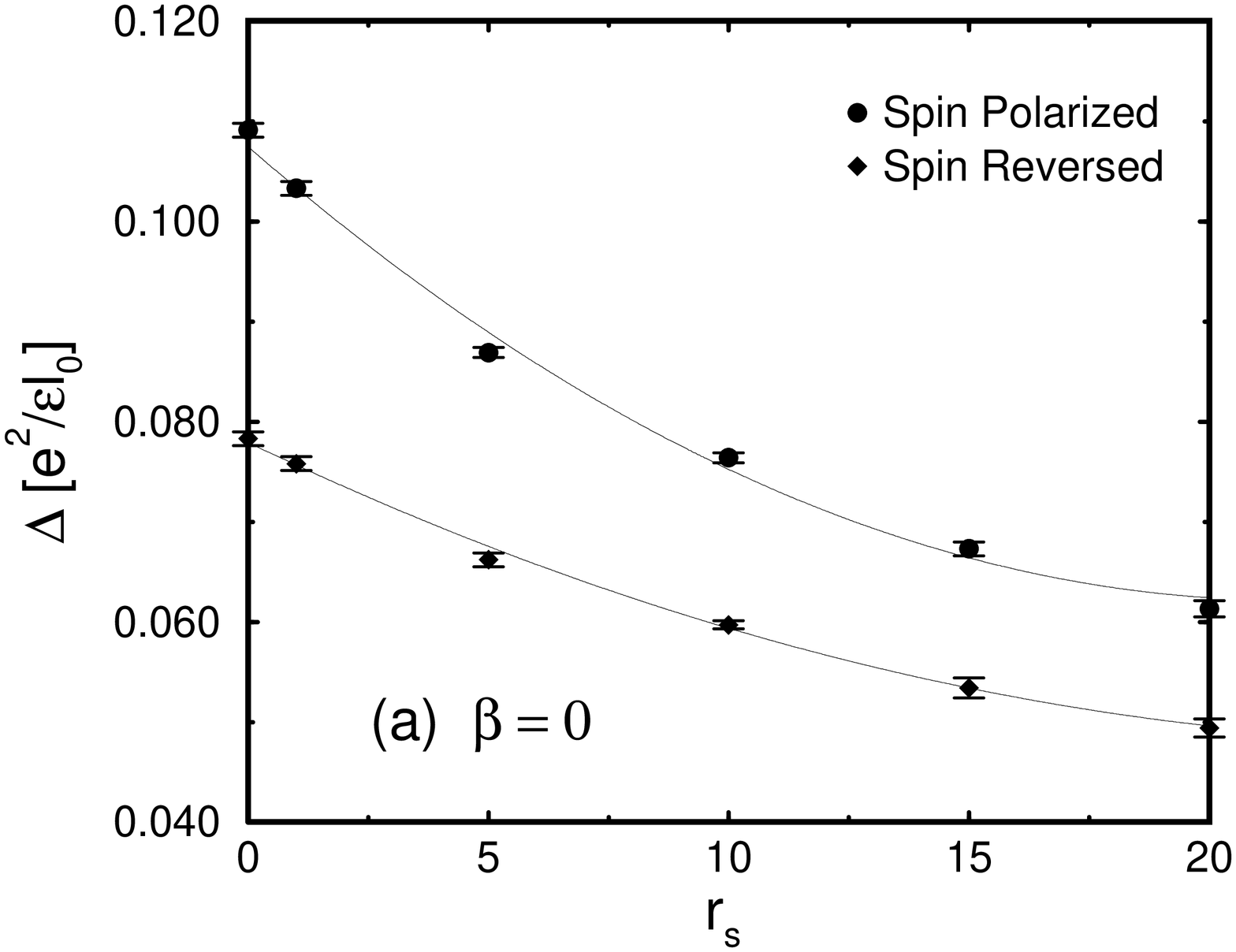,height=3.7in,angle=-90}}
\centerline{
\psfig{figure=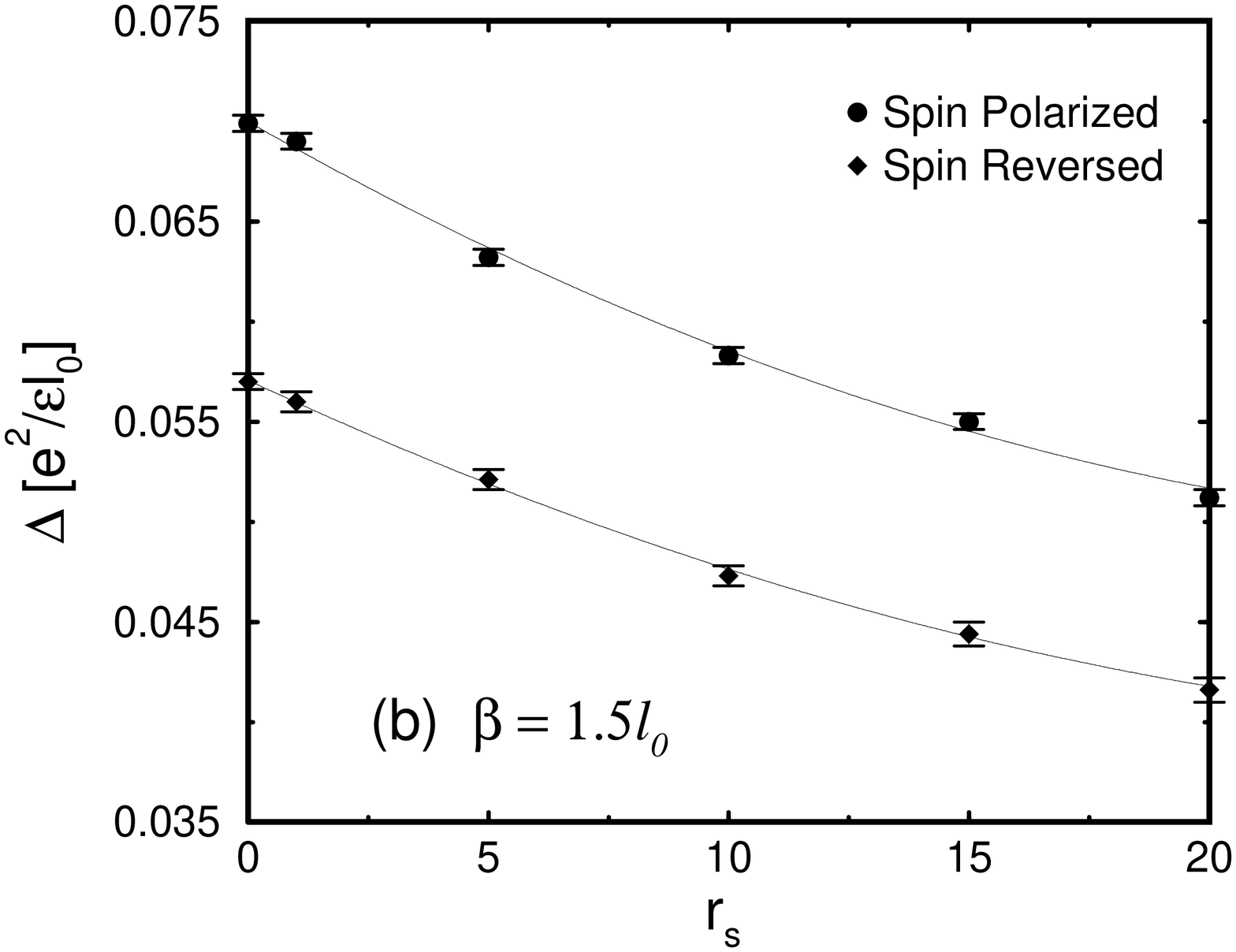,height=3.7in,angle=-90}}
\vskip .5in
\caption{
$\nu=1/3$ energy gaps for creating a quasielectron---quasihole pair at
opposite poles of the sphere vs. $r_s$.  Results are given for both a
spin-polarized (dots) and spin-reversed (diamonds) quasielectron for
thickness parameter {\it (a)} $\beta=0$, and {\it (b)} $\beta = 1.5\
l_0$.  Results are for 20 electrons.  The lines are guides to the
eye.}
\label{gapvsl2}
\end{figure}

\newpage

\begin{figure}
\centerline{
\psfig{figure=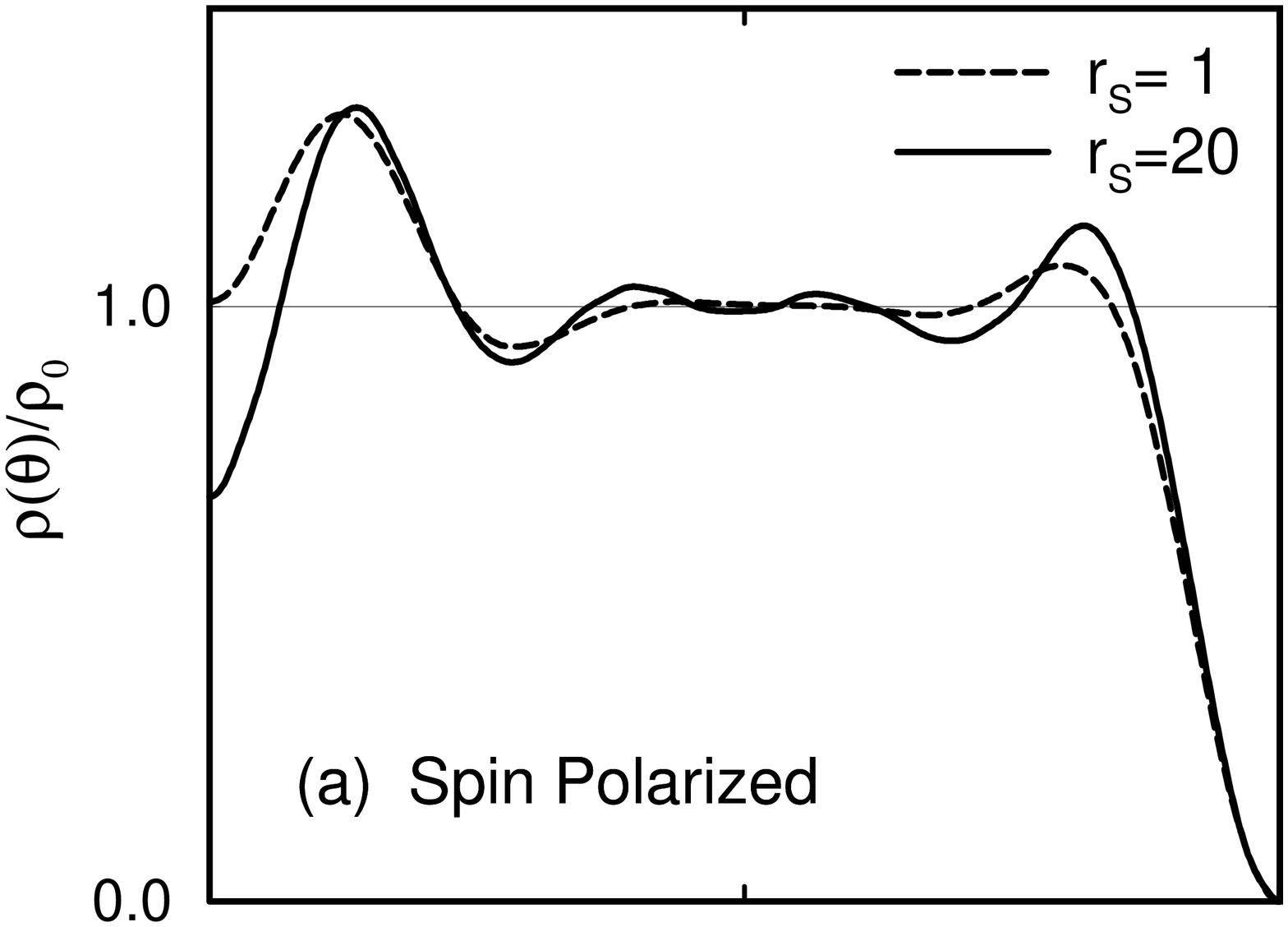,height=3.7in,angle=-90}}
\vskip -.8in
\centerline{
\psfig{figure=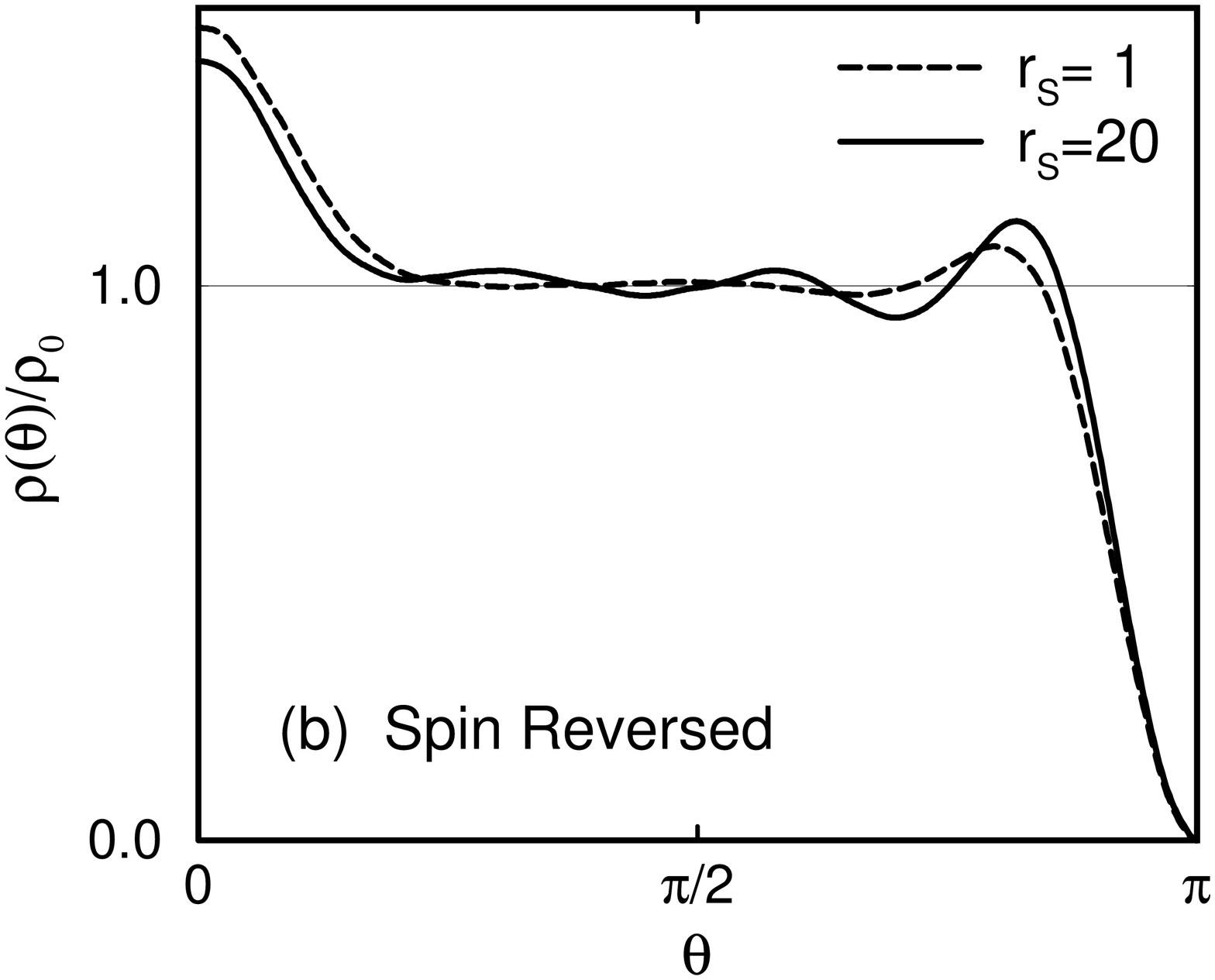,height=3.7in,angle=-90}}
\vskip .5in
\caption{
Mixed estimates of the density profiles of $\nu=1/3$ excited state
wave functions on the sphere with a quasielectron at the top of the
sphere $(\theta = 0)$ and a quasihole on the bottom of the sphere
$(\theta = \pi)$ with $r_s = 1$ (dashed line) and $r_s = 20$ (solid
line), for {\it (a)} the spin-polarized excited state, and {\it (b)}
the spin-reversed excited state.  Results are for 20 electrons.}
\label{density}
\end{figure}

%\end{multicols}


\begin{references}
\bibitem{note1}
To reduce finite size effects we have followed \cite{morf} and quoted
energies in units of $e^2/\epsilon l^\prime_0$ where $l^\prime_0 =
(N+1/N)^{1/2} l_0$ is a rescaled magnetic length.
\end{references}
\end{document}